\documentclass[runningheads]{llncs}
\usepackage{graphicx}
\usepackage[utf8]{inputenc}
\usepackage[T1]{fontenc}
\usepackage{amsmath,amssymb,amsfonts}
\usepackage{listings}
\usepackage{color}
\usepackage{float}
\usepackage{tikz, tkz-base, tkz-fct, pgfplots, pgfplotstable}
\usepackage{siunitx}

\definecolor{mygreen}{rgb}{0,0.6,0}
\definecolor{mygray}{rgb}{0.5,0.5,0.5}
\definecolor{mymauve}{rgb}{0.58,0,0.82}

\begin{document}

\title{GPU Support for Automatic Generation of Finite-Differences Stencil Kernels}
%
%
\author{
  Vitor Hugo Mickus Rodrigues\inst{1}\thanks{The author gratefully acknowledge support from Shell Brasil through the "Novos Métodos de Exploração Sísmica por Inversão Completa das Formas de Onda" project at the Universidade Federal do Rio Grande do Norte, and the strategic importance of the support given by ANP through the R\&D levy regulation.
    This research was supported by the High Performance Computing Center at UFRN (NPAD/UFRN).}\and
  Lucas Cavalcante\inst{1}\and
  Maelso Bruno Pereira\inst{1}\and
  Fabio Luporini\inst{2}\and
  István	Reguly\inst{3}\and
  Gerard	Gorman\inst{2}\and
  Samuel Xavier de Souza\inst{1}
}

\authorrunning{Mickus-Rodrigues et al.}
%
\institute{
  Universidade Federal do Rio Grande do Norte, Natal - RN, Brazil\and
  Imperial College London, London, United Kingdon\and
  Pazmany Peter Catholic University, Budapest, Hungary
}
\maketitle              

\begin{abstract}
    The growth of data to be processed in the Oil \& Gas industry matches the requirements imposed by evolving algorithms based on stencil computations, such as Full Waveform Inversion and Reverse Time Migration. Graphical processing units (GPUs) are an attractive architectural target for stencil computations because of its high degree of data parallelism. However, the rapid architectural and technological progression makes it difficult for even the most proficient programmers to remain up-to-date with the technological advances at a micro-architectural level. In this work, we present an extension for an open source compiler designed to produce highly optimized finite difference kernels for use in inversion methods named Devito\textsuperscript{\textcopyright}. We embed it with the Oxford Parallel Domain Specific Language (OP-DSL) in order to enable automatic code generation for GPU architectures from a high-level representation. We aim to enable users coding in a symbolic representation level to effortlessly get their implementations leveraged by the processing capacities of GPU architectures. The implemented backend is evaluated on a NVIDIA\textsuperscript{\textregistered} GTX Titan Z, and on a NVIDIA\textsuperscript{\textregistered} Tesla V100 in terms of operational intensity through the roof-line model for varying space-order discretization levels of 3D acoustic isotropic wave propagation stencil kernels with and without symbolic optimizations. It achieves approximately 63\% of V100's peak performance and 24\% of Titan Z's peak performance for stencil kernels over grids with 256\textsuperscript{3} points. Our study reveals that improving memory usage should be the most efficient strategy for leveraging the performance of the implemented solution on the evaluated architectures.

\end{abstract}

\section{Introduction}

A wide variety of physical phenomena can be formalized in terms of partial differential equations (PDE) such as sound, heat, diffusion, electrostatics, electrodynamics, fluid dynamics, elasticity, and quantum mechanics. The development of computationally efficient methods for obtaining numerical solutions of PDEs through stencil kernels has been mentioned as a key computational science and engineering challenge to be addressed as one of the "seven dwarfs of computation" for at least the next decade, in 2009 \cite{Williams2009}. In fact, large-scale PDE inversion algorithms that can be solved by finite-difference (FD) schemes used in exploration seismology such as full waveform inversion (FWI) and reverse time migration (RTM) constitute some of the current most computationally demanding problems in industrial and academic research.

In general, a stencil on structured grids is defined as a function that updates a point based on the values of its neighbors. The stencil structure remains constant as it moves from one point in space to the next. In the context of a wave-equation solver, the stencil is described by the support (grid-locations) and the coefficients of FD schemes. Using parallel designs such as graphics processing units (GPU) has relatively recently become the preferred choice to improve existing code for the current commercial and scientific community that performs stencil computations.

However, a significant barrier that has become increasingly more notable is the difficulty in programming these systems. As the hardware architectures grow in complexity, exploiting the potential of these devices requires higher know-how on parallel programming. The issue has further been compounded by a rapidly changing hardware design space, with a wide range of parallel architectures. For example, some designs offer many simple processors vs. fewer complex processors, some depend on multi-threading, and some even replace caches with explicitly addressed local stores. As no conventional wisdom has yet emerged, it is unsustainable for domain scientists to re-write their applications for each new type of architecture regarded that developing and validating a PDE solver usually takes decades of effort.

To address the problem of algorithm sustainability, taking into account the uncertainty in future architectures, one solution involves decoupling the work of a domain scientist and a computer scientist. In this approach, Domain Specific Languages (DSL) are developed by high-performance computing (HPC) specialists, and the specifics of the problem and the numerical solution method are specified in the DSL by the domain scientist. Using source-to-source translation, the numerical solver can be targeted towards different hardware backends. This ensures that only the backend that interfaces with the new architectures need to be written and supported by the translator. The underlying implementation of the solver remains the same, thereby introducing a separation of concerns that results in a direct payoff in productivity.

Interest in building generic DSLs for solving PDEs is not new with early attempts dating back as far as 1970 \cite{Cardenas1970,Cook1988,VanEngelen2003}. More recently, two prominent finite element software packages, FEniCS \cite{Logg2012} and Firedrake \cite{Rathgeber2015}, have demonstrated the power of symbolic computation using the DSL paradigm. The optimization of regular grid and stencil computations has also produced a vast range of libraries and DSLs that aim to ease the efficient automated creation of high-performance codes \cite{Hawick2013,Henretty2013,Membarth2012,Zhang2012}.

In this work, we present an implementation for automatic GPU code-generation to Devito. This objective can be translated into extending Devito’s backend in such a way that the generated stencils are compatible with this target architecture. Currently, two backends exist in Devito: the default backend to run it on standard CPU architectures; and an alternative backend using the YASK stencil compiler to generate optimized C++ code for Intel\textsuperscript{\textregistered} Xeon\textsuperscript{\textregistered} and Intel\textsuperscript{\textregistered} Xeon Phi\textsuperscript{\texttrademark} architectures \cite{Luporini2018}. Our strategy is to utilize one of the Oxford Parallel Domain Specific Languages (OP-DSL), called OPS, to build a third backend for Devito. OPS is a programming abstraction embedded in C/C++ for writing multi-block structured mesh algorithms, and it is composed by the corresponding software library (an Application Programming Interface – API) and code translation tools (compilers) to enable automatic parallelization of the intermediary-level code produced (here, by Devito) using different parallel programming approaches.

As a result, it is expected that executable artifacts wrote in CUDA, OpenACC, OpenCL, OpenMP, and MPI get automatically and transparently composed for a diverse range of hardware from high-level symbolic descriptions of PDEs. It has been shown that OPS generated code is capable of matching or outperforming hand-coded and tuned implementations \cite{Reguly2014}, which implies considerable confidence in such an approach being capable of delivering high performance, code maintainability and future proofing.

It is possible to speculate that it would take much longer not only to compose complex FD problems but also to produce their various hand-coded parallel implementations, each of which would have to be then debugged and validated. The authors claim that the time savings on combining code generation with automatic parallel implementation for state-of-the-art hardware will have a significant impact on the efforts for modeling seismic inversion algorithms.

The remaining of this paper is organized as follows. In Section \ref{sec:Background} we present both the Devito and the OPS compiler, altogether with the model for isotropic wave propagation considered in our study. Section \ref{sec:Methodology} describes how the code generated by Devito should be modified in order to match the syntax of the OPS compiler, and also the roof-line model for evaluating the performance of the generated kernels on the GPU devices considered in this work. In Section \ref{sec:Results} we show and comment on the the results. Section \ref{sec:Conclusion} encloses this paper with concluding remarks.

\section{Background}
\label{sec:Background}
\subsection{Devito}

Devito is a tool to solve partial differential equations (PDEs) which is a mathematical tool to describe numerous  problems that are heavily constrained by physical laws. Some areas in which it has uses are: geophysics, earth and climate science, material science, chemical and mechanical engineering, medical imaging and physics, even in economics. It uses a domain specific language (DSL) as method to simplify development process for the user, and also solve it using finite difference method that is a numerical method.

Devito automatically generates C/C++ code with different levels of optimization for finite-difference schemes from a symbolic Python representation of partial differential equations, with a performance that is competitive with, and often better than, hand-optimized implementations. To illustrate this, consider the Equation \ref{eqn:wave-propagation} that is a wave propagation with a source injection and its initial conditions.

\begin{equation}
\label{eqn:wave-propagation}
\left\{
\begin{aligned}
m(x,y,z)\frac{d^{2}u(x,y,z,t)}{dt^2} - \nabla^2u(x,y,z,t) = q_s,\\
u(x,y,z,0) = 0, \\
\frac{du(x,t)}{dt}|_{t=0} = 0,
\end{aligned}\right.
\end{equation} 
\vspace{0.2cm}

where:
\begin{itemize}
    \item[\textbullet] $m(x,y,z) = \frac{1}{c(x,y,z)^2}$, represents the square slowness model as a function of the three space coordinates $(x,y,z)$;
    \item[\textbullet] $u(t,x,y,z)$, is the spatially varying acoustic wave field in each time step;
    \item[\textbullet] $q_s$, is the source term representing the source injection;
    
\end{itemize}

As Devito uses Sympy library for an easier symbolic representation, writing this equation is as simple as shown in Algorithm \ref{lst:devito-python-code}, which represents a small part of the solution.

\vspace{0.2cm}
\begin{lstlisting}[language=python, caption={Example of Devito declaring an acoustic wave propagation},label={lst:devito-python-code}]
from sympy import Eq, solve
from devito import Function, TimeFunction, Grid

grid = Grid(shape=(size,size))
u = TimeFunction(name='u', grid=grid, space_order=6, time_order=2)
m = Function(name='m', grid=grid)

#Symbolic representation
eqn = Eq(m * u.dt2 - u.laplace)

stencil = solve(eqn, u.forward)[0]
\end{lstlisting}
\vspace{0.2cm}

Devito performs just-in-time compilation and execution, so the domain expert can focus on the mathematical formulations, instead of writing low-level code. Following the example, the C code automatically generated from Devito using Python can be seen in Algorithm \ref{lst:devito-generated-c-code}.

\vspace{0.2cm}
\begin{lstlisting}[language=c, caption={Devito auto generated C code using core backend. Represents the propagation update for stencil of space order 2.},label={lst:devito-generated-c-code}, basicstyle=\scriptsize]
for (int x = x_m; x <= x_M; x += 1)
  {
    #pragma omp simd aligned(damp,m,u:32)
    for (int y = y_m; y <= y_M; y += 1)
    {
      float r0 = 1.0F*dt*m[x+2][y+2][z+2] + 
                 5.0e-1F*(dt*dt)*damp[x+1][y+1][z+1];
      
      u[t1][x+2][y+2][z+2] = 
        1.0F*(-dt*m[x+2][y+2][z+2]*u[t2][x+2][y+2][z+2]/r0 + 
              (dt*dt*dt)*u[t0][x + 1][y + 2][z + 2]/r0 + 
              (dt*dt*dt)*u[t0][x + 2][y + 1][z + 2]/r0 + 
              (dt*dt*dt)*u[t0][x + 2][y + 2][z + 1]/r0 + 
              (dt*dt*dt)*u[t0][x + 2][y + 2][z + 3]/r0 + 
              (dt*dt*dt)*u[t0][x + 2][y + 3][z + 2]/r0 + 
              (dt*dt*dt)*u[t0][x + 3][y + 2][z + 2]/r0) + 
        2.0F*dt*m[x+2][y+2][z+2]*u[t0][x+2][y+2][z+2]/r0 + 
        5.0e-1F*(dt*dt)*damp[x+1][y+1][z+1]*u[t2][x+2][y+2][z+2]/r0 - 
        6.0F*dt*dt*dt*u[t0][x + 2][y + 2][z + 2]/r0;
      }
    }

\end{lstlisting}

The user doesn't even need to see this code, it will all be handled by Devito's compiler and the result from its execution will be available for the developer. Programming the Algorithm \ref{lst:devito-python-code} is much simpler and maintainable than Algorithm \ref{lst:devito-generated-c-code} and it enables the code execution in different architectures using the same python code. 

In this work, we leveraged Devito to support the OPS library (described in Subsection \ref{ss:ops}) for computing stencil kernels in a GPU environment using the CUDA parallel computing platform.

\subsection{OPS}
\label{ss:ops}

OPS provides high-level code abstraction aimed at multi-block suctured grid computations. It can be embedded in C/C++ and its API provides a basic structure for grid computations such as: blocks, datasets defined on these blocks representing constants and state variables, and parallel loops across a block, accessing data defined on the grid points. Which are used to deliver code for different parallel architectures: MPI, OpenMP, OpenACC, CUDA and OpenCL.

\begin{figure}[h]
    \centering
    \includegraphics[width=1\linewidth]{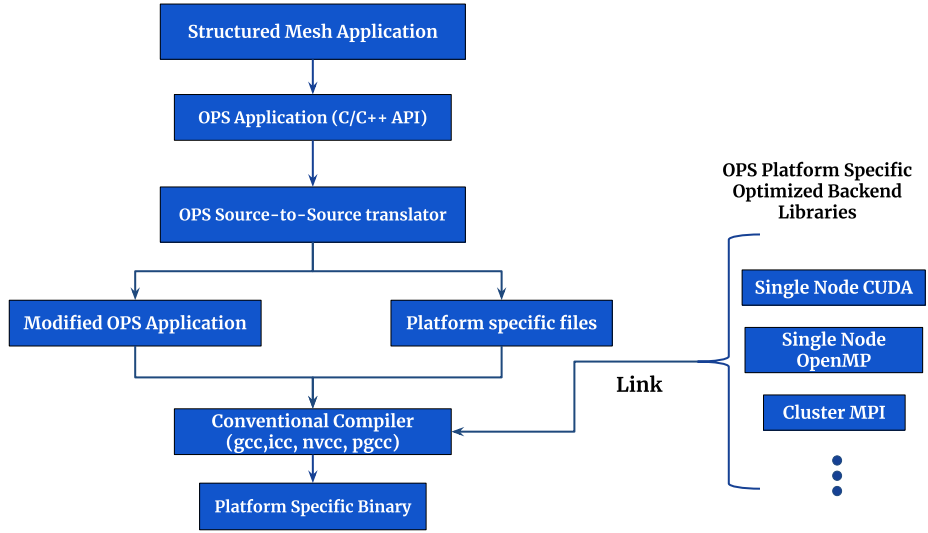}
    \caption{OPS traditional work flow}
    \label{fig:ops-workflow}
\end{figure}

The diagram in Figure \ref{fig:ops-workflow} shows the traditional work flow of OPS programs: starting from the desired structured mesh application then programming the C/C++ algorithm using OPS API, compiling and linking it with OPS libraries and executing the desired platform.

OPS and Devito integration enables automatic code generation for GPU architectures from a high level representation.

\subsection{Devito-OPS Integration}
\label{sec:Methodology}

To accomplish Devito and OPS integration we need to understand the process Devito uses to generate C/C++ code. Devito generates an intermediate representation to perform a sequence of operations to the expressions and iterations, this includes:
\begin{itemize}
    \item[\textbullet] Equations lowering;
    \item[\textbullet] Local analysis;
    \item[\textbullet] Clustering;
    \item[\textbullet] Symbolic optimization;
    \item[\textbullet] Iteration/expression tree (IET) construction;
    \item[\textbullet] Synthesis;
    \item[\textbullet] Operator specialization through backends;
\end{itemize}
\noindent
the last step is where Devito will specialize data types aiming an interested API, which is OPS in this research. Devito with OPS backend share all the compilation pipeline until the specialization.

In this section, we stress seven (i-vii) essential building blocks required to accomplish our prototype solution.

The integration starts with generating \emph{OPS Expression}'s (i), which are expressions translated into OPS syntax. An expression that initially is represented in C/C++ language as 
\begin{verbatim}
    u[t+1][x][y] = u[t][x][y] + 1
\end{verbatim}
has an OPS representation syntax given by:
\begin{verbatim}
    ut10[OPS_ACC0(0,0)] = ut00[OPS_ACC1(0,0)] + 1    
\end{verbatim}
 The array access \texttt{u} in the first representation will be replaced for \texttt{ut10} when indicating a one position forward in the time dimension, and replaced for \texttt{ut00} when accessing the current time dimension. The term \texttt{OPS\_ACC\#(0,0)} is a macro that OPS syntax uses when translating the index to the desired architecture. 
 
 Producing this transformation in Devito requires that the parts of a given expression are separated into nodes. For example, an \emph{Indexed} object containing the indices that corresponds to displacements over dimensions at Devito level, corresponds to C-arrays. The Algorithm \ref{lst:make_ops_ast} illustrates the recursive method used to transform Devito expressions. 

\begin{lstlisting}[language=python, caption={Method to evaluate the given expression and translate to OPS syntax},label={lst:make_ops_ast}]
def make_ops_ast(expr, nfops):
    if expr.is_Symbol or expr.is_Number:
        return expr
    elif expr.is_Indexed:
        return nfops.new_ops_arg(expr)
    else:
        return expr.func(*[make_ops_ast(i, nfops) for i in expr.args])
\end{lstlisting}

We are interested in transforming expressions from offloadable loops. These expressions can be parallelized into a device code that will efficiently get executed by GPU architectures. Parallelizable expressions of the same nest can be grouped inside an outlined function that we call \emph{OPS User Kernel} (ii), called by \emph{ops\_par\_loop} (iii) in the OPS API syntax. The \emph{iteration range} defines the range in which a \emph{OPS User Kernel} will operate over the mesh. It is described as an integer array that defines the boundaries in each spatial dimension. The mesh that will be written into or read from throughout the kernel operation is the dataset that is represented by \emph{ops\_dat} (iv) in the OPS API syntax.

Others API calls needed to generate a compilable OPS code ultimately are:
\begin{itemize}
    \item[(v)] \emph{ops\_init} and \emph{ops\_end} are calls that will mark the beginning and ending of OPS syntax usage. All OPS declarations must be located between these two calls.
    \item[(vi)] \emph{ops\_block} is used to group datasets together.
    \item[(vii)] \emph{ops\_partition} triggers a multi-block partitioning across a distributed memory set of processes.
\end{itemize}

\begin{figure}[h]
\label{fig:devito-ops-integration}
\centering
\includegraphics[width=1\linewidth]{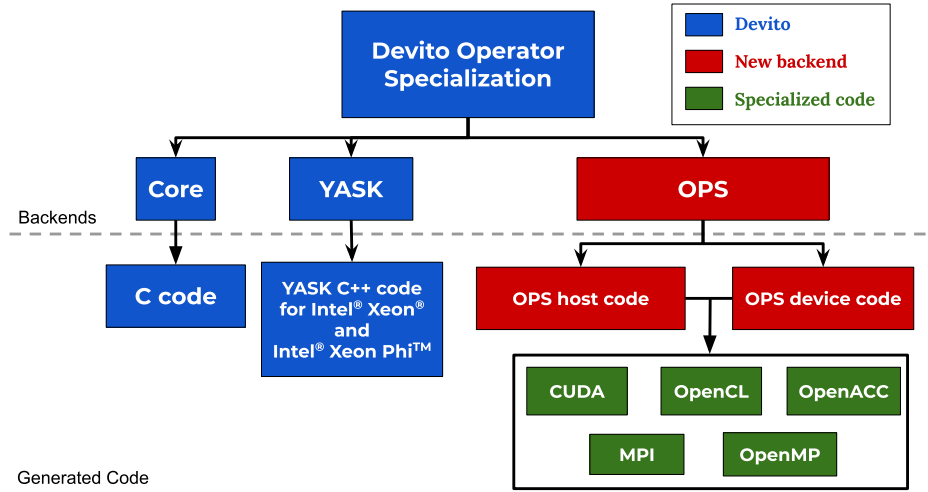}
\caption{Diagram of Devito and OPS integration}
\end{figure}

The diagram in Figure \ref{fig:devito-ops-integration} represents an overview of the Devito and OPS integration. 

The main contribution of this work is a prototype solution that will automatically generate kernel code for a GPU environment. This code can be coupled in a manually generated C code, that is capable of calling this generated kernels. In Section \ref{sec:Conclusion} we discuss in future works how to fully generate the host code.

\section{Experiment}
\label{sec:Experiment}
\subsection{Acoustic Wave Propagation}

The current investigation involved generating isotropic 3D wave propagation stencil kernels in an automatic fashion for two NVIDIA architectures and analyzing the performance of the generated algorithms by the roofline model \cite{Williams2009}. Kepler and Volta were selected as target GPU architectures, with specifications summarized in Table \ref{tab:card_specs}. They use CUDA cores, which is compatible with the syntax supported by the OPS programming interface. Executable artifacts were produced by \textit{NVCC} compiler with flags \texttt{-Xcompiler="-std=c99" -O3}, altogether with specific micro architectural flag depending on the specific architecture.

\begin{table}[]
    \centering
    \begin{tabular}{|l|c|c|}
        \hline
        & Titan Z & Tesla V100 \\
        \hline
        Memory Bandwidth (GB/s) & 336 x2 & 900 \\  
        Single Precision Peak Performance (GFLOPS) & 4746 & 14000 \\
        Double Precision Peak Performance (GFLOPS) & 1582 & 7000 \\
        Memory (GB) & 6 x2 & 16 \\
        \hline
    \end{tabular}
    \caption{Specifications of evaluated graphical cards.}
    \label{tab:card_specs}
\end{table}

\textbf{GTX Titan Z} is a graphics card launched in 2014. It combines two graphics processors for increased performance, although here we only consider one of those cores. This card uses Kepler microarchitecture and specific compilation flags \texttt{-gencode arch=compute\_35, code=sm\_35}. 

\textbf{Tesla V100} is a PCIe 16GB launched in 2017. The micro architectural flag specific for Tesla is \texttt{-gencode arch=compute\_70,code=sm\_70}. 

The performance of the produced solutions was analyzed in terms of their floating-point performance, operational intensity and memory performance through the roofline model. This model reveals the rate between the extent of performance usage and the theoretical peak performance of the evaluated devices.

The maximum performance of each architecture was calculated using Equation \ref{eqn:attainable-peak-performance} considering the hardware specifications described in Table \ref{tab:card_specs}. Any algorithm running in the same architecture will be bound to this very same roof.

\begin{equation}
    \scriptsize
    \parbox{8em}{Attainable\\Peak Performance}
    \text{[GFLOP/s] = min}
    \begin{cases}
        \text{Peak Floating-Point Performance}\\    
        \text{Peak Memory Bandwidth x OI}
\end{cases}
\label{eqn:attainable-peak-performance}
\end{equation}

The Operational Intensity (OI) measures the Dynamic Random Access Memory (DRAM) bandwidth needed by a kernel in a particular architecture.  In the devices considered in this paper, each read or write transaction between the DRAM and the caches have a 32 bytes size. Using this definition, the Equation \ref{eqn:operational-intensity} is used to determine the OI.

\begin{equation}
    \scriptsize
    \text{OI} [FLOP/Byte] = \frac{\text{\# Single Precision Floating-Point Operations}}{(\text{\# Memory Transactions})*32}
    \label{eqn:operational-intensity}
\end{equation}

A kernel performance measures the number of floating-point operations per second. Performance can be directly calculated using Equation \ref{eqn:performance}.

\begin{equation}
    \scriptsize
    \text{Performance} [FLOP/s] = \frac{\text{\# Single Precision Floating-Point Operations}}{\text{Kernel Execution Time}}
    \label{eqn:performance}
\end{equation}

\section{Results}
\label{sec:Results}

Data obtained in previous studies indicated that Devito is able to efficiently utilise  Intel architectures\footnote{Intel\textsuperscript{\textregistered} Xeon\textsuperscript{\textregistered} E5-2690v2 with 10 physical cores, and Intel\textsuperscript{\textregistered} Xeon\textsuperscript{\textregistered} Phi\textsuperscript{TM} accelerator card.} with a high degree of efficiency, while maintaining the ability to increase accuracy by switching to higher order stencil discretization dynamically \cite{Luporini2018}. Luporini et al. show that remarkable speed-ups from 3x up to 4x can be attainable for those architectures on scenarios with what they call "aggressive" optimizations to avoid redundant computation over 3D grids with space order discretization levels varying from 4 to 16. In our study, we measure the performance of a new backend for Devito on the NVIDIA\textsuperscript{\textregistered} architectures GTX Titan Z\textsuperscript{TM} and Tesla V100\textsuperscript{TM} considering scenarios with no symbolic optimizations (basic DSE), and with an aggressive symbolic optimization implemented by Devito (aggressive DSE). An isotropic acoustic wave propagation model with absorbing boundaries as described by Equation \ref{eqn:wave-propagation} is utilized.

In this study, we measured the rate between attainable performance and the peak machine performance according to specifications, for both the considered devices. We take into account the roofline model described in Section \ref{sec:Experiment} to evaluate how efficiently the generated algorithms utilize the GPU for varying space order levels of the generated propagation stencil kernels. For each of the considered space orders we profiled the propagation kernel using \textit{nvprof} \footnote{The \textit{nvprof} profiling tool enables you to collect and view profiling data from the command-line, and is present in the NVIDIA\textsuperscript{\textregistered} CUDA\textsuperscript{\textregistered} Toolkit.} in order to obtain: (a) the number of single precision floating-point operations, (b) the number of memory transactions, and (c) the kernel execution time. 

For each space order, the produced stencil kernel ran five times for 30.000 time steps. Table \ref{tab:perf-titanz} shows the values collected for GTX Titan Z and Table \ref{tab:perf-v100} shows the values collected for V100, for basic and aggressive symbolic optimization levels, and space orders levels of 8, 12, 16 and 24. The values for OI are obtained according to Equation \ref{eqn:operational-intensity} whereas the values for performance are obtained according to Equation \ref{eqn:performance}. 

\begin{figure}[h!]
    \includegraphics[width=1\linewidth]{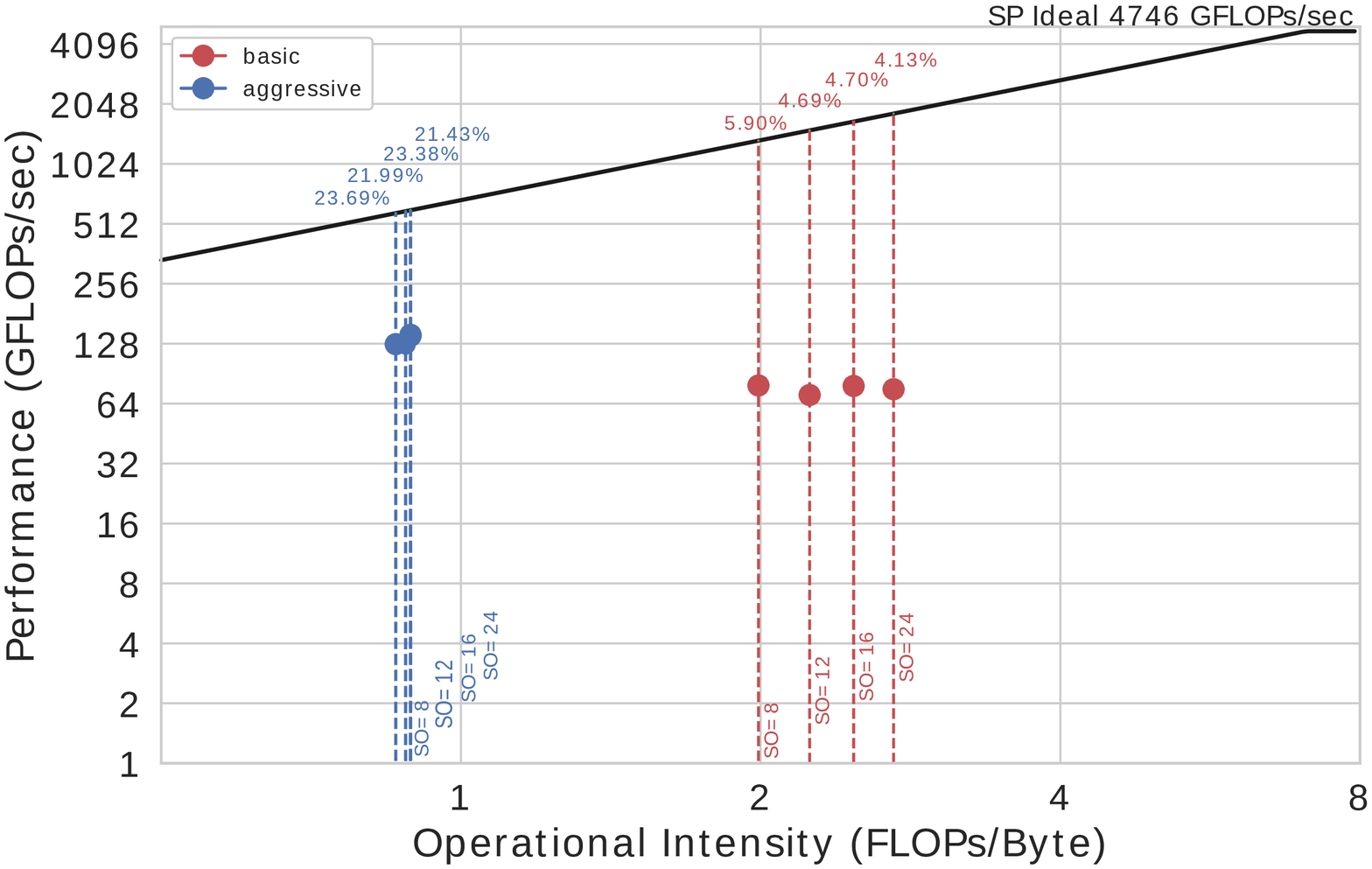}
    \caption{Roofline chart for GTX Titan Z GPU. Propagation field with 256$^3$ points and space order values of 8, 12, 16 and 24 using Devito optimizations aggressive and basic.}
    \label{fig:titanz-roofline}
\end{figure}

\begin{figure}[h!]
    \includegraphics[width=1\linewidth]{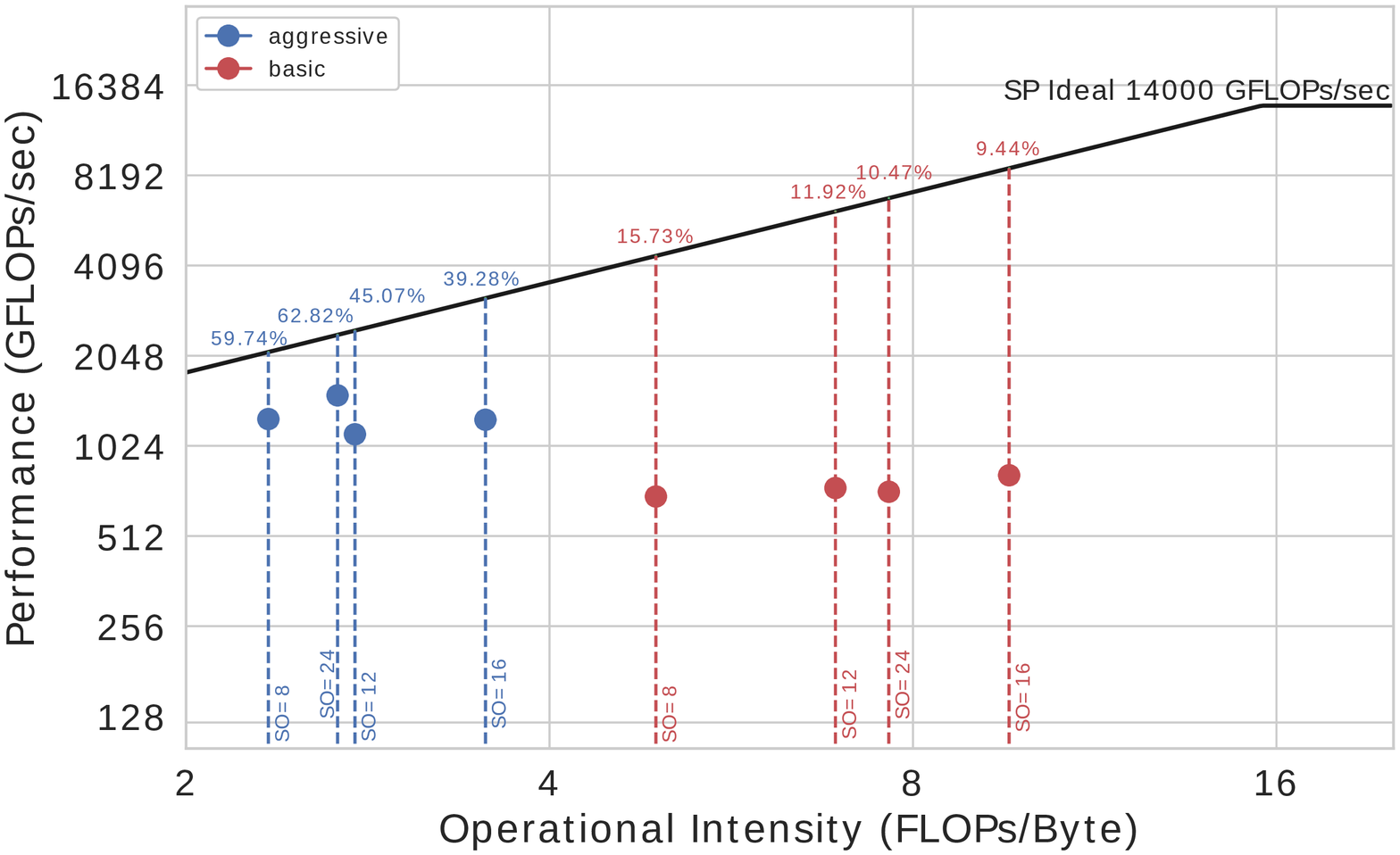}
    \caption{Roofline chart for V100 GPU. Propagation field with 256$^3$ points and space order values of 8, 12, 16 and 24 using Devito optimizations aggressive and basic.}
    \label{fig:v100-roofline}
\end{figure}

Figures \ref{fig:titanz-roofline} and \ref{fig:v100-roofline} display the OI (FLOP/Byte) versus performance (GFLOP/s) from the values found in Tables \ref{tab:perf-titanz} and \ref{tab:perf-v100}, respectively. Each of the points in those plots are characterized by two values: (i) the space order, and (ii) the percentage from the device peak performance.
The performance bounds were obtained from vendor peak performance specifications in Table \ref{tab:card_specs}.

\begin{table}[]
\centering
\begin{tabular}{|cccc|cc|}
\hline
\textbf{\begin{tabular}[c]{@{}c@{}}Space\\ Order\end{tabular}} & \textbf{\begin{tabular}[c]{@{}c@{}}FP 32 \\ Count\end{tabular}} & \textbf{\begin{tabular}[c]{@{}c@{}}Memory \\ Operations\end{tabular}} & \textbf{\begin{tabular}[c]{@{}c@{}}Execution \\ Time (s)\end{tabular}} & \textbf{\begin{tabular}[c]{@{}c@{}}OI \\ (Flop/Byte)\end{tabular}} & \textbf{\begin{tabular}[c]{@{}c@{}}Performance \\ (GFlop/s)\end{tabular}} \\ \hline
\multicolumn{6}{|c|}{\textit{Basic Optimization}} \\ \hline
8 & 1,450,112,268 & 22,722,746 & 553.92 & \textbf{1.99} & \textbf{78.54} \\
12 & 2,013,392,118 & 28,068,109 & 854.39 & \textbf{2.24} & \textbf{70.70} \\
16 & 2,375,372,938 & 29,871,728 & 907.72 & \textbf{2.48} & \textbf{78.51} \\
24 & 2,898,342,158 & 33,348,001 & 1,150.01 & \textbf{2.71} & \textbf{75.61} \\ \hline
\multicolumn{6}{|c|}{\textit{Aggressive Optimization}} \\ \hline
8  & 641,887,345 & 22,637,047 & 135,73 & \textbf{0.89} & \textbf{141.88} \\
12 & 760,134,906 & 27,737,029 & 179.15 & \textbf{0.86}  &  \textbf{127.29} \\
16 & 842,931,505 & 29,704,549 & 180,55 & \textbf{0.89} & \textbf{140.06} \\
24 & 929,761,776 & 32,926,331 & 219,76 & \textbf{0.88} & \textbf{126.92} \\ \hline
\end{tabular}
\vspace{0.2cm}
\caption{Data collected from profiling propagation kernel in GTX Titan Z using \textit{nvprof}.}
\label{tab:perf-titanz}
\end{table}

\begin{table}[]
\centering
\begin{tabular}{|cccc|cc|}
\hline
\textbf{\begin{tabular}[c]{@{}c@{}}Space\\ Order\end{tabular}} & \textbf{\begin{tabular}[c]{@{}c@{}}FP 32 \\ Count\end{tabular}} & \textbf{\begin{tabular}[c]{@{}c@{}}Memory \\ Operations\end{tabular}} & \textbf{\begin{tabular}[c]{@{}c@{}}Execution \\ Time (s)\end{tabular}} & \textbf{\begin{tabular}[c]{@{}c@{}}OI \\ (Flop/Byte)\end{tabular}} & \textbf{\begin{tabular}[c]{@{}c@{}}Performance \\ (GFlop/s)\end{tabular}} \\ \hline
\multicolumn{6}{|c|}{\textit{Basic Optimization}} \\ \hline
8 & 1,450,996,129 & 9,245,436 & 553.92 & \textbf{4.90} & \textbf{693.77} \\
12 & 2,013,446,796 & 9,112,947 & 854.39 & \textbf{6.90} & \textbf{740.48} \\
16 & 2,375,384,531 & 7,722,032 & 907.72 & \textbf{9.61} & \textbf{816.86} \\
24 & 2,898,311,328 & 11,862,338 & 1,150.01 & \textbf{7.64} & \textbf{719.60} \\ \hline
\multicolumn{6}{|c|}{\textit{Aggressive Optimization}} \\ \hline
8  & 641,882,304 & 9,256,098  & 15.31 & \textbf{2.18} & \textbf{1258.16} \\
12 & 760,133,342 & 9,289,727  & 20.37 & \textbf{2.56} & \textbf{1119.42} \\
16 & 842,930,745 & 8,026,245  & 20.21 & \textbf{3.28} & \textbf{1251.51} \\
24 & 929,760,267 & 11,670,483 & 18.48 & \textbf{2.49} & \textbf{1509.60} \\ \hline
\end{tabular}
\vspace{0.2cm}
\caption{Data collected from profiling propagation kernel in V100 using  \textit{nvprof}.}
\label{tab:perf-v100}
\end{table}

Considering the results for GTX Titan Z in Figure \ref{fig:titanz-roofline}, we can see that the operation intensity increases with higher space order levels for basic optimization, whereas the operation intensity are nearly the same for an aggressive optimization.

Considering the results for GTX Titan Z in Figure \ref{fig:titanz-roofline}, we can see that for the basic optimization, the operation intensity increase with higher the space orders, while using the aggressive optimization they almost did not differ. One can also see that aggressive optimization produces code with better performance than with basic optimization in all scenarios, enabling approximately 24\% of peak performance to be achieved versus 6\% for the basic scenario.

Executing the experiment in the V100 graphic card, we achieve better performance, as illustrated in Figure \ref{fig:v100-roofline}. Performance gains using aggressive optimization goes from approximately 16\% to 63\%. It is worth noting that there is a decrease in OI for \texttt{so} 24, this result was not expected as there are more operations in higher \texttt{so}. Looking at the data from Table \ref{tab:perf-v100} we can verify that the amount of data transferred in \texttt{so} 24 is 45\% higher than the \texttt{so} 16, while the difference in data transfer in the other scenarios was at most 15\%. We can thus conclude that the amount of data needed for \texttt{so} 24 is much larger than expected, which indicates that memory accesses in GPU are not coalesced for this case.

In both GTX Titan Z and V100 tests, the aggressive optimization led to three times higher peak performance than the basic optimization. The results from the aggressive optimization corroborate results presented in a related experiment, Luporini et al. \cite{Luporini2018} that enabled Devito to generate code for the YASK framework and obtained peak performances going from 53\% to 63\% for Intel\textsuperscript{\textregistered} Xeon\textsuperscript{\textregistered} and Intel\textsuperscript{\textregistered} Xeon\textsuperscript{\textregistered} Phi\textsuperscript{\texttrademark} architectures.

Another analysis possible due to the Roofline model is for future optimizations. All the points are located before the ridge point at both the roofline plots in Figures \ref{fig:titanz-roofline} and \ref{fig:v100-roofline}, which indicates that all the tested cases are memory bounded instead of compute bounded. This means that the produced codes should get greater benefits from optimizations targeted to perform memory exchanges more efficiently than from optimizations focused on increasing throughput. Therefore, enabling FLOPs reduction and data locality such as common sub-expression elimination, factorization, and code motion should be considered as a priority for future works.

\section{Conclusion}
\label{sec:Conclusion}

The open-source project Devito\textsuperscript{\textregistered} \cite{Louboutin2018,Luporini2018} has been attracting the attention of academic \cite{Mojica2019,Witte2019} and industrial \cite{Yount2017} community. As a DSL for seismic inversion applications, it already provides a set of automated performance optimizations during code generation that allow user applications to fully utilize the target hardware without changing the model specification, such as vectorization, shared-memory parallelism, loop blocking, auto-tuning, common sub-expression elimination (CSE), cross-iteration redundancy elimination (CIRE), expression hoisting and factorization. Devito also supports distributed-memory parallelism via MPI, and several halo-exchange schemes are available. Classic optimizations such as computation-communication overlap (relying on asynchronous progress engine) are implemented. It can be integrated with a wide variety of methods (e.g. L-BFGS-B\footnote{Large-scale Bound-constrained Optimization
}) for solving minimization problems, such as in FWI. It can perform FWI on distributed memory parallel computers with Dask. It also implements support for standard CPU architectures, and for Intel\textsuperscript{\textregistered} Xeon\textsuperscript{\textregistered} and Intel\textsuperscript{\textregistered} Xeon Phi\textsuperscript{\texttrademark} architectures. However, the support to code specialization for GPU architectures is yet a work in progress.

In this study, we created an extension of Devito to enable code generation for the OPS syntax.  We also evaluated the new backend in terms of processor performance concerning off-chip memory traffic for varying space order discretization levels on the NVIDIA\textsuperscript{\textregistered} devices GTX Titan Z\textsuperscript{TM} and Tesla V100\textsuperscript{TM}. We found that the implemented backend achieves up to 62.82\% of the peak performance on V100, which is consistent with results from work using Devito to generate YASK framework code \cite{Luporini2018}. We also observed that isotropic 3D wave propagation stencil kernels generated with aggressive symbolic optimizations have three times higher peak performance than with no symbolic optimizations. This study, therefore, indicates that it is possible to use the available power of GPU architectures in Devito for solving seismic inversion algorithms.

This work is the first study to our knowledge that investigates a seamless coupling between Devito and OPS compilers. However, some limitations are worth noting as the capability of the implemented solution still only covers source injection and forward propagation. The forward model is the basis for further implementations of inversion processes using Devito operators. Yet, in order to enable a seamless source-to-source translation of FWI algorithms, future work should provide support for receiver interpolation and backward propagation as well. Moreover, the automatic generation of host code, responsible for calling the device code that will execute in GPU, is currently in implementation. Finally, to complete Devito integration, it is necessary to automatically translate, compile, and execute the GPU code through the Devito pipeline and return the result from the execution to the Devito workflow.



\end{document}